\documentclass[aps,prl]{revtex4}

\usepackage{graphicx}
\usepackage{dcolumn}
\usepackage{bm}
\begin{document}
\addtolength{\headheight}{1.5cm}
\title{
Time-dependent Nonlinear Optical Susceptibility of an \\
Out-of-Equilibrium Soft Material}
\author{
Neda Ghofraniha$^{1}$, Claudio Conti$^{2}$, Giancarlo
Ruocco$^{1,2}$ and Francesco Zamponi$^{3}$}

\affiliation{
$^{1}$Dipartimento di Fisica, Universita' di Roma ``La
Sapienza'', P. A. Moro 2, 00185, Roma, Italy\\
$^{2}$SOFT-INFM-CNR, c/o Universita' di Roma ``La Sapienza'',  P.
A. Moro 2, 00185, Roma, Italy\\
$^{3}$Laboratoire de Physique Th\'eorique, \'Ecole Normale
Sup\'erieure,
24 Rue Lhomond, 75231 Paris Cedex 05, France
}

\date{\today}

\begin{abstract}
We investigate the time-dependent nonlinear
optical absorption of a clay dispersion (Laponite) in 
organic dye (Rhodamine B) water solution displaying 
liquid-arrested state transition. Specifically, we
determine the characteristic time $\tau_D$ of the nonlinear susceptibility
build-up due as to the Soret effect.
By comparing $\tau_D$
with the relaxation time provided by standard 
dynamic light scattering measurements we report on 
the decoupling of the two collective diffusion times at the 
two very different length
scales during the aging of the out-of-equilibrium system.
With this demonstration experiment we also show the potentiality of
nonlinear optics measurements in the study of the late stage 
of arrest in soft materials.
\end{abstract}


\maketitle

Light can influence matter microscopic properties providing
an optical response that depends on the laser power. 
The nonlinear optical susceptibility of a wide 
variety of materials (e.g. simple liquids~\cite{Nag03}, 
magnetic fluids~\cite{Alv03,Sog07}, liquid crystals~\cite{KhooBook},
polymeric thin films~\cite{Lia05} and
nano-composite layers~\cite{delcoso04})
has been investigated
by using different techniques.
Time-resolved Z-scan~\cite{Oli94} has been
used to distinguish between different nonlinear effects in crystals
and in liquids and transient holographic 
gratings~\cite{Spi00} allowed to study
the thermal diffusion of nanospheres.
Nonlinear optics in soft material has been 
explored in diluted colloidal suspensions~\cite{Ash82a},
in metallic nanoparticles~\cite{Gom07OL}
with different chemical stabilizers~\cite{Gom07JOSAB},
in presence of nanoparticles aggregation~\cite{Wan05} and
recently in out-of-equilibrium systems~\cite{Con06,Gho07}. 

Complex fluids far from thermodynamic equilibrium 
are known to display {\it aging}~\cite{Bou97,Leticia}, meaning that
dynamical quantities such as correlations and responses
depend on the time spent since the sample preparation
({\it waiting time} $t_w$).
The same is expected for
the dynamics of nonlinear optical self-action effects, like self-phase
modulation and nonlinear absorption, 
an issue that so far remains unexplored.
These effects can be used to determine  the Soret coefficient
and the collective diffusion coefficient of the colloidal
particles moving on length scales much larger than particles size.
Therefore the presented 
nonlinear optics experiments can be very useful to characterize
dynamical properties of colloidal systems, for instance
{\it i)} to study the dynamics of jammed systems characterized by long 
range interactions and {\it ii)} to analyze the dynamical properties
of weakly turbid samples, not detectable by standard light scattering measurements.

In this Letter we report on the time-dependent optical nonlinearity 
of a complex out-of-equilibrium system during its dynamics 
slowing down.
The investigated soft-medium is a dye-doped clay (Laponite) dispersed in water.
The suspension is prepared by solving the Laponite powder, supplied by Laporte Ltd., 
in a solution of Rhodamine-B (RhB) and  deionized water at 
0.1 mM concentration. 
The suspension at 1.6 wt\% clay
concentration  is then stirred vigorously until it  is cleared
and filtered through $0.22\;\mu$m pore size Millipore filter. 
The sample is prepared in air and we
take the time when
the suspension is filtered as the starting aging 
time ($t_w=0$). 
Once dispersed in water Laponite platelets 
start to aggregate and their dynamics slows down
leading to a liquid-arrested state transition
~\cite{Ruz04PRL}. 
In the mentioned RhB-water-Laponite suspension 
all dye molecules attach to the clay particles surface~\cite{Arb98aLang}. 
We estimate about 10 dye
molecules per platelet, much less than the maximum allowed adsorption sites
(about 500 as provided by Laporte Ltd. data-sheet).
The RhB molecules
absorb visible light that locally heats up the medium and 
the induced thermal gradient influences
both water density and dielectric properties 
(thermal effect)~\cite{Sin00JAP}
and colloidal particle concentration 
(thermo-diffusive or Soret effect)~\cite{Rus04aJOSA}. 
Both phenomena lead to a nonlinear optical response
characterized by distinct time scales:
$\sim ms$ and $\sim s$, respectively. The thermal effect is responsible
of local refraction changes (thermal lens), while the Soret effect in principle
can produce either absorption or refraction variations. 
To study this twofold effect 
we perform both non-linear refraction and absorption
measurements.

By using the Z-scan approach, for refraction experiments the light source is 
a CW pumped diode laser operating at 
wavelength $\lambda=532\, nm$
modulated  by a mechanical shutter
with opening time $700\mu s$.
The beam is focused by a 75
mm focal-length lens providing a 35 $\mu$m beam waist radius $w$
at the sample position.  A photo-detector with rise-time $14\, ns$
and an angular acceptance of $0.73^\circ$
was used to probe the light power.
Being the
beam spot on the detector larger than the photo-diode surface, only the light
transmitted along the beam axis is collected, giving access 
to the nonlinear phase shift~\cite{Bah90a}.
For absorption experiments we use the same setup in addition
with a $25\,mm$ focal-length lens before the detector, 
thus collecting the whole transmitted beam.
The colloidal dispersion is syringed in a glass cuvette with
transverse dimensions $10 \times 35\,mm^2$ 
and $1\,mm$ thickness along the beam axis.
Each single measurement is the response of a different point
of the sample not previously illuminated.
We investigate the doped clay response at four
different incident laser powers for each $t_w$, 
during which we consider unchanged the dynamics and structural 
properties of Laponite at 1.6\% w/w, 
as inferred
from Dynamic Light Scattering (DLS) measurements on the same sample.

\begin{figure}
\begin{center}
\includegraphics[width=7.5cm]{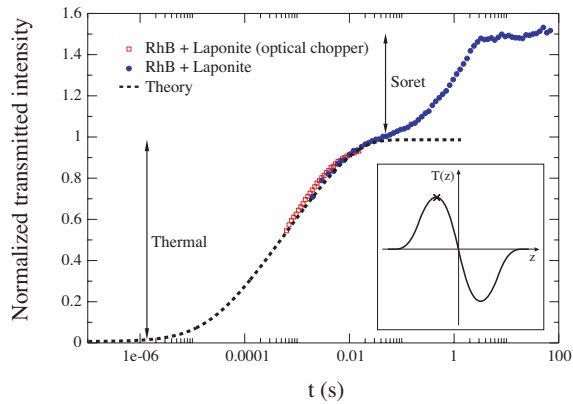}
\caption{{\bf Refraction}. Normalized transmitted light intensity of 
0.1mM RhB-water+1.6\% w/w Laponite dispersion
compared to the theoretical thermal lens curve (solid line)- see text. 
Input power 8.3 mW.
Inset: sketch of  z-scan profile; the cross corresponds to the sample
position.} 
\label{fig1}
\end{center}
\end{figure}

Refraction measurements 
are aimed to
distinguish between thermal and Soret 
time-scales.
In Fig.~\ref{fig1} the normalized transmitted light intensity
is reported displaying  the two distinct
times.
The intensity is normalized such that
its value gives $\Delta \bar n(t)/\Delta \bar n_T$, where $\Delta \bar n(t)$ is 
the nonlinear phase shift per unit length and unit
intensity (radial integral of the bell shaped $\Delta \bar n(r,t)$, 
with $r$ the transversal radial coordinate)
and $\Delta \bar n_T$ corresponds 
to the saturation of the thermal contribution.
At short times the temperature gradient 
induces mostly water density variation, which provides a 
change of the index of refraction $\Delta\, n (t)<0$ (thermal lens)
that broadens the input beam. 
Being the sample  positioned before the focus of the lens (z=0,
as shown in the inset of~\ref{fig1}), the transmitted beam
gets narrower augmenting the detected light intensity.
In Fig.~\ref{fig1} 
the intensity growth at large times 
is due to the dyed clay thermal-diffusion,
that induces variation of light absorption influencing the refractivity.
Z-scans experiments~\cite{Gho07} show that 
the RhB-platelets diffusion drastically alters
the nonlinear absorption that is considered with details
in the following.
In Fig.~\ref{fig1}, at short times, the thermal effect is
compared to the theoretical curve from~\cite{Wu90JAP},
calculated with our experimental parameters. From the comparison 
we have $t^\star=65\,ms$ as the time
when the thermal gradient can be regarded as stationary.

\begin{figure}
\begin{center}
\includegraphics [width=7.5cm, height=8.5cm]{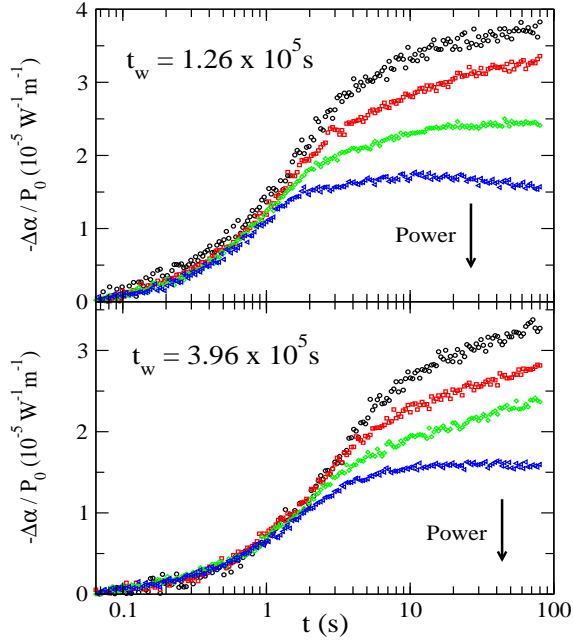}
\caption{{\bf Absorption}. Scaled absorption coefficient of  
0.1mM RhB-water+1.6\% w/w Laponite dispersion
at two different aging times. The four inject power values are: 
8.3 mW, 11 mW, 14 mW and 17.8 mW.} 
\label{fig2}
\end{center}
\end{figure}

In the following  
nonlinear optical absorption and its evolution during
the gelation  of the soft material is considered.
The transmitted 
intensity is $I(r,t)=I_0(r)\,\text{exp}[-\alpha(r,t)\,L]$ with
$I_0(r)=I_0\,\text{exp}(-2r^2/w^2)$ the input,
$w$ the beam waist, $I_0$ the peak value
and $L$
the sample thickness.
The absorption coefficient  
$\alpha(r,t)=\alpha_0+\Delta\alpha[I(r,t)]$ can be written as the sum
of a constant term (the linear absorption coefficient)
and a time and laser intensity dependent term $\Delta\alpha[I(r,t)]$
that is related  to the gradient of particles mass fraction 
$\Delta c[I(r,t)]$ 
by
\begin{equation}\label{deltaalpha}
\Delta \alpha[I(r,t)]=\left(\frac{\partial \alpha}
{\partial c}\right)_{\bar c}\,\Delta c[I(r,t)],
\end{equation}
where $\bar c$ is the average mass fraction.
$\Delta\alpha[I(r,t)]$ is the response of
the out-of-equilibrium 
system to the solicitation $I(r,t)$,
which is proportional to the local 
temperature gradient responsible of the platelets diffusion.
We consider the series expansion 
of $\Delta\alpha[I(r,t)]$ in terms of $I_0$:
the first term is the susceptibility linear in theperturbation
and it can 
be described by the linear hydrodynamics equations
[equations~(\ref{heat}) and~(\ref{massflux}) below].
The nonlinear absorption coefficient  $\Delta\alpha[I(r,t)]$ 
can be obtained in our experiments by writing
$\Delta\alpha(r,t)=-[I(r,t)-I(r,t=t^\star)]/L\,I(r,t=t^\star)$,
where $t^\star$
is the starting time of the colloid  Soret effect, 
$I(r,t^\star)=I_0(r)\,\text{exp}[-\alpha(r,t^\star)\,L]$
and we retain the first term in the expansion
$\text{exp}[-\Delta\alpha(r,t)\,L]$.
In this way we examine the contribution to the
nonlinear optical susceptibility $\Delta\alpha(t)$, that is only
due to platelets motion and hence  its dependence on
the structural properties of the jelling system.
In the following we scale down the 
response function $\Delta\alpha(r,t)$
to the input power $P_0$ as $\Delta\alpha(r,t)/P_0$. 
In Fig.~\ref{fig2} we show the output signal
$S(t)=-\langle \Delta\alpha(r,t)\rangle_r\,/P_0$
at four powers and two different aging times
($\langle \Delta \alpha(r,t)\rangle_r=
\int_0^\infty {\mbox d} r\,r\,e^{\frac{-2r^2}{w^2}}\,\Delta \alpha(r,t)/
\int_0^\infty {\mbox d} r\,r\,e^{\frac{-2r^2}{w^2}}$ is the radial average over the incident 
intensity profile).
In the range $65\,ms<t<1\,s$
the signals at different input powers collapse on one curve defining
the linear response regime.   
For $t>1\,s$ higher orders terms are relevant; these terms 
give direct access to the dynamical heterogeneity length scales~\cite{Bou05} 
and their characterization
will be reported elsewhere.

\begin{figure}
\begin{center}
\includegraphics[width=7.5cm] {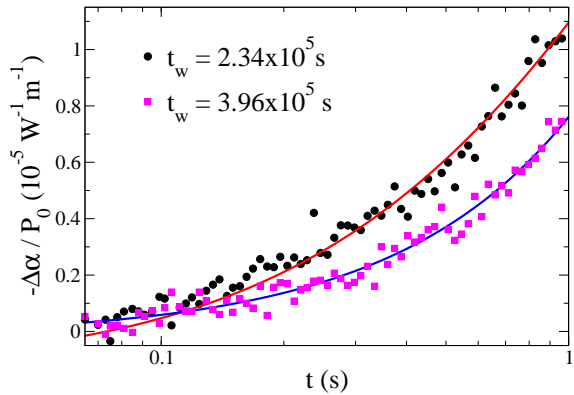}
\caption{{\bf Absorption}. Scaled absorption coefficient of    
0.1mM RhB-water+1.6\% w/w Laponite dispersion
and the fitting curves (full lines) by using expression~(\ref{radial}).} 
\label{fig3}
\end{center}
\end{figure}

\begin{figure}
\begin{center}
\includegraphics[width=7.5cm,height=8.5cm] {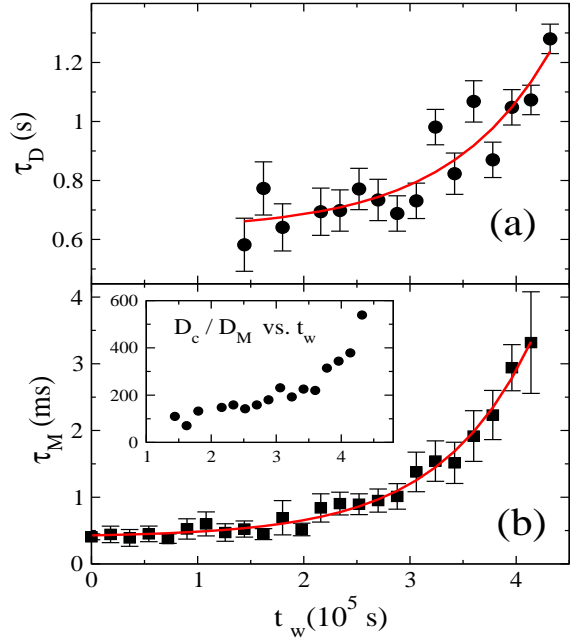}
\caption{Characteristic thermo-diffusion time (a)
and mean relaxation time (b)
vs. $t_w$ and the
fitting exponential curves (full lines). 
Inset: the ratio $D_c / D_M$ vs. $t_w$} 
\label{fig4}
\end{center}
\end{figure}

In the ``linear'' regime ( $65\,ms<t<1\,s$) the induced particle
mass fraction change $\Delta c[I(r,t)]$
is given by the solution of the heat-mass flux  coupled equations
~\cite{deGBook}
\begin{eqnarray} \label{heat}
\frac{\partial}{\partial t} \Delta T(r,t) 
& = & D_T \nabla^2 \Delta T(r,t) \\ \nonumber
& + & \frac{D_T\,D_D\,\bar T\,\mu_c}{k}\nabla^2 \Delta c(r,t)
+\frac{D_T\,\alpha_0}{k}\, I(r) \\ \label{massflux}
\frac{\partial}{\partial t}\Delta c(r,t)  
& = & D_c \nabla^2 \Delta c(r,t) \\ \nonumber
& + & D_c S_T \bar c (1-\bar c) \nabla^2 \Delta T(r,t)  
\end{eqnarray}
where $\Delta T(r,t)$
is the radial temperature variation due to the local 
light absorption, $D_T$  and $k$ are  water thermal 
diffusivity  and conductivity respectively, $\bar T$ is the 
bulk temperature $\mu_c$ is the the variation of
colloid's chemical potential in presence of a
concentration gradient and $D_D$ is the
Dufour coefficient;
$D_c$ is the particles collective diffusion coefficient and $S_T$ is
the Soret coefficient.
As shown in Fig.~\ref{fig1} clay concentration gradient builds up 
once the temperature profile reaches a stationary
state, this allows to
neglect the Dufour effect in the heat equation and 
to take $\nabla^2 \Delta T(r,t)$
in~(\ref{massflux}) from the stationary
solution of~(\ref{heat}).
In decoupling the two effects we consider 
negligible the 
Soret feedback on the temperature profile, being 
$(|\alpha_2|\,I_0)/\alpha_0 \sim 10^{-5}$,
where the
coefficients $\alpha_0=0.86\,mm^{-1}$ and $\alpha_2=-2.8\, nm/W$
are obtained by fitting
$\log[I_0/I(t=1s)]=L\,(\alpha_0+\alpha_2\,I_0)$ vs. $I_0$.
In this approximation
the colloidal mass fraction variation
leads to
\begin{eqnarray}\label{deltac}
\Delta c(r,t) & = & \frac{\bar c\,S_T\,\alpha_0\,P_0}{4\pi\,k}
\left[Ei\left(-\frac{2\,r^2}{w^2}\right)\right. \\ \nonumber
              & - & \left.Ei\left(-\frac{2\,r^2}{w^2}\frac{1}{1+2t/\tau_D}\right)\right],
\end{eqnarray}
being  $P_0=I_0\pi w^2$ the beam power,
$\tau_D=w^2/4 D_c$ the characteristic diffusion time
over the beam spot size, and $Ei$ the exponential
integral function.
To analyze the observed nonlinear absorption  in our experiments by means
of equation~(\ref{deltac}) we consider 
the output signal 
\begin{eqnarray}\label{radial}
S(t) & = & -\langle \Delta\alpha(r,t)\rangle_r\,/P_0 \\ \nonumber
     & = & L\left(\frac{\partial \alpha}{\partial c}\right)_{\bar c}\,
\frac{\bar c\,S_T\,\alpha_0}{4\pi\,k}\log\left(1+\frac{t}{\tau_D}\right)
\end{eqnarray}
as obtained by using expressions~ (\ref{deltaalpha}) and (\ref{deltac}).
We use equation~(\ref{radial}) to fit 
the experimental results at different $t_w$ values.
We show in Fig.~\ref{fig3} the output data with
the fitting  curves (solid lines) at two aging times 
and in Fig.~\ref{fig4}-a
the fitted characteristic thermo-diffusion
times $\tau_D$ at various $t_w$. 
We find that this characteristic time exponentially
grows  during aging as the structural relaxation time 
$\tau_M(t_w)$ obtained from DLS
measurements, which are contemporarily performed on the same sample.
DLS data
were fitted assuming a correlation function made by 
square of the sum of an
exponential function with relaxation time $\tau_1$ and a stretched
exponential function with relaxation time $\tau_2$ and stretching
coefficient $\beta$. The mean relaxation time is defined as $\tau_M
=\tau_2\,{\beta^{-1}}\,\Gamma\,\left({\beta^{-1}} \right)$
with $\Gamma(x)$ the usual Euler gamma function.
Indeed we fit both $\tau_D(t_w)$ and the mean relaxation time
$\tau_M(t_w)$ 
(representing the slow dynamics of the correlation function)
with the exponential growth law 
$\tau=\tau_0+\tau_1\,{\mbox exp}(\mu\,t_w)$ 
(full lines trough data)
as shown in Fig.~\ref{fig4}-a and Fig.~\ref{fig4}-b respectively. 
Both $\tau_D$ and $\tau_M$ describe the collective diffusion of
interacting particles on different length scales: DLS probes dynamics
on $1/q\simeq 50\, nm$ and nonlinear susceptibility on $w\simeq 35\,\mu m$.
Although we find comparable growth rates $\mu$ 
from the fitting procedure,
the ratio $D_c/D_M$ [with $D_c=w^2/(4 \, \tau_D)$ and $D_M=1/(q^2 \, \tau_M)$]
reported
in the inset of~\ref{fig4}-b 
as a function of $t_w$ increases of approximately one order of magnitude in the
accessible waiting-time window.
Moreover at very low $t_w$ the two diffusion coefficients are not the same.
This behavior can be attributed to the long ranged (screened electrostatic) 
interactions between Laponite particles,
characterized by the peculiarity of non- homogeneous surface
charge distribution, that makes the system strongly interacting
on relatively large length scales even at very small~$t_w$~\cite{Ruz08}.
Another possible explanation might be 
the deviation to the Fluctuation-Dissipation Relation 
(FDR) in an out-of-equilibrium glassy system~\cite{Bou97}.
Indeed the diffusion constant measured in optical absorption is related 
to the response of density
to an external field, while the one measured in DLS is related 
to the correlation of density fluctuations.
The mentioned discrepancy could be also related to a
supradiffusive regime as shown in~\cite{Sch06}.

We finally estimate the Soret coefficient 
as $S_T\simeq 0.02\,K^{-1}$ from 
the amplitude of the fitting expression~(\ref{radial}),
by considering 
($\frac{\partial \alpha}{\partial c})_{\bar c}\simeq\alpha_0/\bar c$,
and by using water thermal conductivity $k=0.58\,W/m\,K$.
Notably enough the Soret coefficient, defined as
$S_T=-(1/\bar c)\,{\mbox d}[\Delta c(t_w)]/{\mbox d}T$,
is constant during the aging process and $S_T>0$ thus
indicating that the platelets migrate from
hot to cold regions 
being ${\mbox d}(\Delta c(t_w))/{\mbox d}T<0$.

In conclusion we measured the time-dependent 
nonlinear optical susceptibility
of a soft-material evolving toward structural arrest.
We characterized the dynamics of the nonlinear absorption
in terms of material parameters  
as the characteristic diffusion time $\tau_D$
and the Soret coefficient.
We have evidence of decoupling of two different diffusion times probed
on two distinct length scales signaling  the existence of long ranged
interactions between the colloidal particles and suggesting 
the violation of the FDR~\cite{Bel01}.
The nonlinear optical response can be hence used to
investigate dynamical properties of colloidal gels characterized by
long range interactions, especially the presented
method can cover length ranges much larger than particles size, 
not detectable by scattering measurements; moreover it can
be really useful in analyzing the dynamics of weakly turbid samples,
not easily investigated by standard light scattering techniques.

We thank Giovanni Bolle for helping us in the experimental set-up
realization and Francesco Sciortino for useful discussions.
C.C. acknowledges the ERC Grant No.201766.


\begin{thebibliography}{29}
\expandafter\ifx\csname
natexlab\endcsname\relax\def\natexlab#1{#1}\fi
\expandafter\ifx\csname bibnamefont\endcsname\relax
  \def\bibnamefont#1{#1}\fi
\expandafter\ifx\csname bibfnamefont\endcsname\relax
  \def\bibfnamefont#1{#1}\fi
\expandafter\ifx\csname citenamefont\endcsname\relax
  \def\citenamefont#1{#1}\fi
\expandafter\ifx\csname url\endcsname\relax
  \def\url#1{\texttt{#1}}\fi
\expandafter\ifx\csname
urlprefix\endcsname\relax\def\urlprefix{URL }\fi
\providecommand{\bibinfo}[2]{#2}
\providecommand{\eprint}[2][]{\url{#2}}


\bibitem[{\citenamefont{Naga Srinivas et~al.}(2003)\citenamefont{Naga Srinivas, Venugopal Rao, and Narayana Rao }}]{Nag03}
\bibinfo{author}{\bibfnamefont{N. K. M.}~\bibnamefont{Naga Srinivas}},
  \bibinfo{author}{\bibfnamefont{S.}~\bibnamefont{Venugopal Rao}},
  \bibnamefont{and}
  \bibinfo{author}{\bibfnamefont{D.}~\bibnamefont{Narayana Rao}},
  \bibinfo{journal}{J. Opt. Soc. Am. B} \textbf{\bibinfo{volume}{20}},
  \bibinfo{pages}{2470} (\bibinfo{year}{2003}). 

\bibitem[{\citenamefont{Alves et~al.}(2003)\citenamefont{Alves, Bourdon and Figueiredo Neto}}]{Alv03}
\bibinfo{author}{\bibfnamefont{S.}~\bibnamefont{Alves}},
  \bibinfo{author}{\bibfnamefont{A.}~\bibnamefont{Boudon}},
  \bibnamefont{and}
  \bibinfo{author}{\bibfnamefont{A.M.}~\bibnamefont{Figueiredo Neto}},
  \bibinfo{journal}{J. Opt. Soc. Am. B} \textbf{\bibinfo{volume}{20}},
  \bibinfo{pages}{713} (\bibinfo{year}{2003}). 

\bibitem[{\citenamefont{Soga et~al.}(2007)\citenamefont{Soga, Alves, Campos, Tourinho, Depeyrot, and Figueiredo Neto}}]{Sog07}
\bibinfo{author}{\bibfnamefont{D.}~\bibnamefont{Soga}}
\bibnamefont{{\it et~al.}},
  \bibinfo{journal}{J. Opt. Soc. Am. B} \textbf{\bibinfo{volume}{24}},
  \bibinfo{pages}{49} (\bibinfo{year}{2007}). 



\bibitem[{\citenamefont{Khoo}(1987)\citenamefont{Khoo}}]{KhooBook}
\bibinfo{author}{\bibfnamefont{I.C.}~\bibnamefont{Khoo}},
  \emph{\bibinfo{title}{Liquid Crystals: Physical Properties and Nonlinear Optical Phenomena}}
  (\bibinfo{publisher}{Wiley}, \bibinfo{address}{New York},
  \bibinfo{year}{1995}).

\bibitem[{\citenamefont{Naga Srinivas et~al.}(2006)\citenamefont{Liang, and Zhou}}]{Lia05}
\bibinfo{author}{\bibfnamefont{J.C.}~\bibnamefont{Liang}},
  \bibnamefont{and}
  \bibinfo{author}{\bibfnamefont{X.Q.}~\bibnamefont{Zhou}},
  \bibinfo{journal}{J. Opt. Soc. Am. B} \textbf{\bibinfo{volume}{22}},
  \bibinfo{pages}{2468} (\bibinfo{year}{2005}).
\bibinfo{author}{\bibfnamefont{V.}~\bibnamefont{Rosso}}
\bibnamefont{{\it et~al.}},
  \bibinfo{journal}{J. Non-Cryst. Solids} \textbf{\bibinfo{volume}{342}},
  \bibinfo{pages}{140} (\bibinfo{year}{2004}).
\bibinfo{author}{\bibfnamefont{A.}~\bibnamefont{Dhinojwala}},
  \bibinfo{author}{\bibfnamefont{J.C..}~\bibnamefont{Hooker}},
  \bibinfo{author}{\bibfnamefont{J.M.}~\bibnamefont{Torkelson}},
  \bibinfo{journal}{J. Non-Cryst. Solids} \textbf{\bibinfo{volume}{172}},
  \bibinfo{pages}{286} (\bibinfo{year}{1994}).

\bibitem[{\citenamefont{del~Coso et~al.}(2004)\citenamefont{del Coso, Requejo-Isidro, Solis, Gonzalo, Afonso}}]{delcoso04}
\bibinfo{author}{\bibfnamefont{R.}~\bibnamefont{del Coso}}
\bibnamefont{{\it et~al.}},
  \bibinfo{journal}{J. Appl. Phys.} \textbf{\bibinfo{volume}{95}},
  \bibinfo{pages}{2755} (\bibinfo{year}{2004}).

\bibitem[{\citenamefont{Oliveira et~al}(1994)\citenamefont{Oliveira, and Zilio}}]{Oli94}
\bibinfo{author}{\bibfnamefont{L.C.} \bibnamefont{Oliveira}}
  \bibnamefont{and} 
  \bibinfo{author}{\bibfnamefont{S.C.}~\bibnamefont{Zilio}},
  \bibinfo{journal}{Appl. Phys. Lett.} \textbf{\bibinfo{volume}{65}},
  \bibinfo{pages}{2121} (\bibinfo{year}{1994}).
\bibinfo{author}{\bibfnamefont{J.}~\bibnamefont{Wang}}
\bibnamefont{{\it et~al.}},
  \bibinfo{journal}{J. Opt. Soc. Am. B} \textbf{\bibinfo{volume}{11}},
  \bibinfo{pages}{1009} (\bibinfo{year}{1994}). 

\bibitem[{\citenamefont{Spill et~al.}(2000)\citenamefont{Spill, K\"{o}hler,
  Lindenblatt, and Schaertl}}]{Spi00}
\bibinfo{author}{\bibfnamefont{R.}~\bibnamefont{Spill}},
 \bibinfo{author}{\bibfnamefont{W.}~\bibnamefont{K\"{o}hler}},
  \bibinfo{author}{\bibfnamefont{G.}~\bibnamefont{Lindenblatt}},
 \bibnamefont{and} \bibinfo{author}{\bibfnamefont{W.}~\bibnamefont{Schaertl}},
  \bibinfo{journal}{Phys. Rev. E} \textbf{\bibinfo{volume}{62}},
  \bibinfo{pages}{8361} (\bibinfo{year}{2000}).
\bibinfo{author}{\bibfnamefont{J.}~\bibnamefont{Rauch}},
  \bibnamefont{and}
  \bibinfo{author}{\bibfnamefont{W.}~\bibnamefont{K\"{o}hler}},
  \bibinfo{journal}{Phys. Rev. Lett.} \textbf{\bibinfo{volume}{88}},
  \bibinfo{pages}{185901-1} (\bibinfo{year}{2002}).

\bibitem[{\citenamefont{A.~Ashkin}(1982)}]{Ash82a}
\bibinfo{author}{\bibfnamefont{P.~W.~S.} \bibnamefont{A.~Ashkin},
  \bibfnamefont{J.~M.Dziedzic}}, \bibinfo{journal}{Opt. Lett.}
  \textbf{\bibinfo{volume}{7}}, \bibinfo{pages}{276} (\bibinfo{year}{1982}).

\bibitem[{\citenamefont{Gomez et~al.}(2007)\citenamefont{Gomez, de Araujo, Rossi, Masunaga, and Jardim}}]{Gom07OL}
\bibinfo{author}{\bibfnamefont{Luis A.}~\bibnamefont{Gomez}}
\bibnamefont{{\it et~al.}}
  \bibinfo{journal}{Opt. Lett.} \textbf{\bibinfo{volume}{32}},
  \bibinfo{pages}{1435} (\bibinfo{year}{2007}).

\bibitem[{\citenamefont{Gomez et~al.}(2007)\citenamefont{Gomez, de Araujo, Brito Silva, and Galembeck}}]{Gom07JOSAB}
\bibinfo{author}{\bibfnamefont{Luis A.}~\bibnamefont{Gomez}}
\bibnamefont{{\it et~al.}},
  \bibinfo{journal}{J. Opt. Soc. Am. B} \textbf{\bibinfo{volume}{24}},
  \bibinfo{pages}{2136} (\bibinfo{year}{2007}).
 
\bibitem[{\citenamefont{Wang Gang et~al.}(2005)\citenamefont{Gang Wang, Yu Zhang, Yiping Cui, Muyun Duan and Mi Liu}}]{Wan05}
\bibinfo{author}{\bibfnamefont{Gang}~\bibnamefont{Wang}}
\bibnamefont{{\it et~al.}},
  \bibinfo{journal}{Opt. Comm.} \textbf{\bibinfo{volume}{249}},
  \bibinfo{pages}{311} (\bibinfo{year}{2005}).

\bibitem[{\citenamefont{Conti et~al}(2006)\citenamefont{Conti, Ghofraniha, Ruocco, and Trillo}}]{Con06}
\bibinfo{author}{\bibfnamefont{C.} \bibnamefont{Conti}},
 \bibinfo{author}{\bibfnamefont{N.} \bibnamefont{Ghofraniha}}, 
  \bibinfo{author}{\bibfnamefont{G.} \bibnamefont{Ruocco}}
  \bibnamefont{and}
  \bibinfo{author}{\bibfnamefont{S.}~\bibnamefont{Trillo}},
  \bibinfo{journal}{Phys. Rev. Lett.} \textbf{\bibinfo{volume}{97}},
  \bibinfo{pages}{123903} (\bibinfo{year}{2006}).

\bibitem[{\citenamefont{Ghofraniha et~al}(2007)\citenamefont{Ghofraniha, Conti, and Ruocco}}]{Gho07}
  \bibinfo{author}{\bibfnamefont{N.} \bibnamefont{Ghofraniha}},
  \bibinfo{author}{\bibfnamefont{C.} \bibnamefont{Conti}},  \bibnamefont{and}
  \bibinfo{author}{\bibfnamefont{G.} \bibnamefont{Ruocco}},
  \bibinfo{journal}{Phys. Rev. B} \textbf{\bibinfo{volume}{75}},
  \bibinfo{pages}{224203} (\bibinfo{year}{2007}).


\bibitem[{\citenamefont{Bouchaud et~al}(1997)\citenamefont{Bouchaud, Cugliandolo, Kurchan,and Mezard}}]{Bou97}
  \bibinfo{author}{\bibfnamefont{J.P.} \bibnamefont{Bouchaud}}
\bibnamefont{{\it et~al.}},
  \bibinfo{journal}{arXiv:cond-mat/9702070v1} (\bibinfo{year}{1997}).


\bibitem[{\citenamefont{Cugliandolo et~al}(2003)\citenamefont{Cugliandolo}}]{Leticia}
  \bibinfo{author}{\bibfnamefont{L.} \bibnamefont{Cugliandolo}}, in
  \bibinfo{author}{\bibfnamefont{J.L.} \bibnamefont{Barrat}}, 
  \bibinfo{author}{\bibfnamefont{M.} \bibnamefont{Feigelman}},
  \bibinfo{author}{\bibfnamefont{J.} \bibnamefont{Kurchan}}, \bibnamefont{and}
  \bibinfo{author}{\bibfnamefont{J.} \bibnamefont{Dalibard eds.}},
  \bibinfo{journal}{Les Houches lecture notes session LXXVII}
  \bibinfo{pages}{367}(\bibinfo{year}{2003}),

\bibitem[{\citenamefont{Ruzicka
  et~al.}(2004{\natexlab{a}})\citenamefont{Ruzicka, Zulian, and
  Ruocco}}]{Ruz04PRL}
\bibinfo{author}{\bibfnamefont{B.}~\bibnamefont{Ruzicka}},
  \bibinfo{author}{\bibfnamefont{L.}~\bibnamefont{Zulian}}, \bibnamefont{and}
  \bibinfo{author}{\bibfnamefont{G.}~\bibnamefont{Ruocco}},
  \bibinfo{journal}{Phys. Rev. Lett} \textbf{\bibinfo{volume}{93}}, \bibinfo{pages}{258301}
  (\bibinfo{year}{2004}{\natexlab{a}}).
\bibinfo{author}{\bibfnamefont{D.}~\bibnamefont{Bonn}}
\bibnamefont{{\it et~al.}},
  \bibinfo{journal}{Europhys. Lett.} \textbf{\bibinfo{volume}{45}},
  \bibinfo{pages}{52} (\bibinfo{year}{1999}).


\bibitem[{\citenamefont{Arbeloa et~al.}(1998)\citenamefont{Arbeloa, Martinez,
  Arbeloa, and Arbeloa}}]{Arb98aLang}
\bibinfo{author}{\bibfnamefont{F.~L.} \bibnamefont{Arbeloa}}
\bibnamefont{{\it et~al.}},
\bibinfo{journal}{Langmuir}\textbf{\bibinfo{volume}{14}},
   \bibinfo{pages}{4566} (\bibinfo{year}{1998}).


\bibitem[{\citenamefont{Sinha et~al.}(2000)\citenamefont{Sinha, Ray, and
  Dasgupta}}]{Sin00JAP}
\bibinfo{author}{\bibfnamefont{S.}~\bibnamefont{Sinha}},
  \bibinfo{author}{\bibfnamefont{A.}~\bibnamefont{Ray}}, \bibnamefont{and}
  \bibinfo{author}{\bibfnamefont{K.}~\bibnamefont{Dasgupta}},
  \bibinfo{journal}{J. App. Phys.} \textbf{\bibinfo{volume}{87}},
  \bibinfo{pages}{3222} (\bibinfo{year}{2000}).


\bibitem[{\citenamefont{Rusconi et~al.}(2004)\citenamefont{Rusconi, Isa, and
  Piazza}}]{Rus04aJOSA}
\bibinfo{author}{\bibfnamefont{R.}~\bibnamefont{Rusconi}},
  \bibinfo{author}{\bibfnamefont{L.}~\bibnamefont{Isa}}, \bibnamefont{and}
  \bibinfo{author}{\bibfnamefont{R.}~\bibnamefont{Piazza}},
  \bibinfo{journal}{J. Opt. Soc. Am. B} \textbf{\bibinfo{volume}{21}},
  \bibinfo{pages}{605} (\bibinfo{year}{2004}).


\bibitem[{\citenamefont{Bahae et~al.}(1990)\citenamefont{Bahae, Said, Wei,
  Hagan, and {van Stryland}}}]{Bah90a}
\bibinfo{author}{\bibfnamefont{M.~S.} \bibnamefont{Bahae}}
\bibnamefont{{\it et~al.}}'
  \bibinfo{journal}{IEEE J. Quantum Electron.} \textbf{\bibinfo{volume}{26}}, \bibinfo{pages}{760}
  (\bibinfo{year}{1990}).

\bibitem[{\citenamefont{Wu and Dovici}(1990)}]{Wu90JAP}
\bibinfo{author}{\bibfnamefont{S.}~\bibnamefont{Wu}} \bibnamefont{and}
  \bibinfo{author}{\bibfnamefont{N.~J.} \bibnamefont{Dovici}},
  \bibinfo{journal}{J. App. Phys.} \textbf{\bibinfo{volume}{67}},
  \bibinfo{pages}{1170} (\bibinfo{year}{1990}).

\bibitem[{\citenamefont{Bouchaud et~al}(2005)\citenamefont{Bouchaud, and Biroli}}]{Bou05}
  \bibinfo{author}{\bibfnamefont{J.P.} \bibnamefont{Bouchaud}},
  \bibnamefont{and}
  \bibinfo{author}{\bibfnamefont{G.} \bibnamefont{Biroli}},
  \bibinfo{journal}{Phys. Rev. B} \textbf{\bibinfo{volume}{72}},
  \bibinfo{pages}{064204} (\bibinfo{year}{2005}).



\bibitem[{\citenamefont{De~Groot}(1962)\citenamefont{De~Groot and Mazur}}]{deGBook}
\bibinfo{author}{\bibfnamefont{S.R.}~\bibnamefont{De~Groot}},
  \bibnamefont{and}
  \bibinfo{author}{\bibfnamefont{P.}~\bibnamefont{Mazur}},
  \emph{\bibinfo{title}{Nonequilibrium Thermodynamics}}
  (\bibinfo{publisher}{(North Holland}, \bibinfo{address}{Amsterdam},
  \bibinfo{year}{1962}).


\bibitem[{\citenamefont{Ruzicka
  et~al.}(2008{\natexlab{a}})\citenamefont{Ruzicka, Zulian, Angelini, Sztucki, 
Moussaid  and Ruocco}}]{Ruz08}
\bibinfo{author}{\bibfnamefont{B.}~\bibnamefont{Ruzicka}}
\bibnamefont{{\it et~al.}},
  \bibinfo{journal}{Phys. Rev. E} \textbf{\bibinfo{volume}{77}}, \bibinfo{pages}{020402(R)}
  (\bibinfo{year}{2008}{\natexlab{a}}).


\bibitem[{\citenamefont{Schosseler
  et~al.}(2008{\natexlab{a}})\citenamefont{Schosseler, Kaloun, Skouri and Munch }}]{Sch06}
\bibinfo{author}{\bibfnamefont{F.}~\bibnamefont{Schosseler}},
  \bibinfo{author}{\bibfnamefont{S.}~\bibnamefont{Kaloun}}, 
 \bibinfo{author}{\bibfnamefont{M.}~\bibnamefont{Skouri}},
\bibnamefont{and}
  \bibinfo{author}{\bibfnamefont{J.~P.}~\bibnamefont{Munch}},
  \bibinfo{journal}{Phys. Rev. E} \textbf{\bibinfo{volume}{73}}, \bibinfo{pages}{021401}
  (\bibinfo{year}{2006}).

\bibitem[{\citenamefont{Bellon et~al}(2001)\citenamefont{Bouchaud, and Biroli}}]{Bel01}
  \bibinfo{author}{\bibfnamefont{L.} \bibnamefont{Bellon}},
 \bibinfo{author}{\bibfnamefont{S.} \bibnamefont{Ciliberto}},
  \bibnamefont{and}
  \bibinfo{author}{\bibfnamefont{C.} \bibnamefont{Laroche}},
  \bibinfo{journal}{Europhys. Lett.} \textbf{\bibinfo{volume}{53}},
  \bibinfo{pages}{511} (\bibinfo{year}{2001})



\end{thebibliography}
\end{document}